\documentclass[aps,prb,onecolumn,floatfix]{revtex4}

\usepackage{epsf}
\usepackage{subfigure}
\usepackage{amsmath}
\usepackage{amssymb}
\usepackage{bm}
\usepackage{graphicx}
\usepackage{epstopdf}
\newcommand{\mean}[1]{\left \langle #1 \right \rangle}

\newcommand{\be}{\begin{equation}}
\newcommand{\ee}{\end{equation}}
\newcommand{\bea}{\begin{eqnarray}}
\newcommand{\eea}{\end{eqnarray}}

\newcommand{\parent}[1]{\left( #1 \right)}

\begin{document}

\title{\bf Nonequilibrium large deviations are determined by equilibrium dynamics}

\author{David Andrieux}

\begin{abstract}
We show that the large deviations of nonequilibrium systems are determined by the fluctuations of associated equilibrium dynamics. 
In particular, this implies that numerical calculations and experimental measurements of nonequilibrium fluctuations can be done at equilibrium. 
\end{abstract}

\maketitle

\vskip 0,25 cm

\section{Example}

We consider a driven random walk on a ring composed of $N$ sites. 
The transition probabilities are $p$ to jump from site $i \rightarrow i+1$ and $1-p$ to jump from $i \rightarrow i-1$. 
The system is at equilibrium when the thermodynamic force $A = \ln \parent{p/(1-p)}$ vanishes and out of equilibrium otherwise.

We consider the fluctuations of the current. The current is measured as $J(n) = \pm 1$ if transition at time $n$ is in the direction $i \rightarrow i \pm 1$. The current fluctuations are characterized by the generating function
\bea
Q_A(\lambda) =\lim_{n \rightarrow \infty} \frac{1}{n} \ln \mean{ \exp \parent{-\lambda \sum_{k=1}^n J(k)} }_A  \, .
\eea
The question we address in this paper is: {\it What do equilibrium fluctuations reveal about nonequilibrium ones? }\\

A straightforward calculation shows that the generating function takes the form
\bea
Q_{{\rm EQ}}  (\lambda) = \ln \parent{e^{-\lambda} + e^{\lambda}} - \ln 2 
\label{Qeq}
\eea
at equilibrium, and 
\bea
Q_A (\lambda) = \ln \parent{e^{-\lambda} + e^{\lambda-A} } - \ln \parent{1+e^{-A}}
\label{QA}
\eea
out of equilibrium. 

The generating functions (\ref{QA}) and (\ref{Qeq}) are related as
\bea
Q_A (\lambda) = Q_{{\rm EQ}} (\lambda - A/2) - Q_{{\rm EQ}} (- A/2) \, .
\eea
This shows that {\bf the nonequilibrium fluctuations can be obtained from the equilibrium ones}.
In the next section we demonstrate that this construction can be generalized to {\it any} nonequilibrium system.

\section{Result}

We consider a Markov chain characterized by a transition matrix $P = \parent{P_{ij}} \in \mathbb{R}^{N\times N}$ on a finite state space.
We assume that the Markov chain is primitive, i.e., there exists an $n_0$ such that $P^{n_0}$ has all positive entries. 

We associate to $P$ an equilibrium dynamics as follows. 
Let $\bar{E}_{ij}[P] = \sqrt{P_{ij}P_{ji}}$ and denote its Perron root by $\chi$ and its right Perron vector by $x$. Then if $D = {\rm diag}(x_1, \ldots, x_N)$ we have that
\begin{eqnarray}
E = \frac{1}{\chi} D^{-1} \ \bar{E} [P] \ D
\label{mapT}
\end{eqnarray}
defines an equilibrium stochastic dynamics. 

We consider the thermodynamic currents (see Ref. \cite{S76} for a definition) and their fluctuations. We have the\\

\newpage

{\bf Theorem.} {\it Let $P$ be a Markov dynamics with affinities $\pmb{A}$ and $E$ its associated equilibrium dynamics (\ref{mapT}). Then their large deviations functions are related as
\bea
Q_P (\pmb{\lambda}) = Q_E (\pmb{\lambda}-\pmb{A}/2) - Q_E (-\pmb{A}/2) \, .
\eea
}\\

DEMONSTRATION:   See Theorem 2 in Ref. \cite{A12b}. $\Box$\\ 

The key insight behind this result is the association between the dynamics $P$ and $E$. 
This link structures the space of stochastic dynamics into equivalence classes (see Ref. \cite{A12b} for a detailed study).\\

This theorem reveals that nonequilibrium large deviations are determined by equilibrium fluctuations. 
In particular, numerical calculations and experimental measurements of nonequilibrium large deviations can be done at equilibrium. 
Thanks to this result, we can use the simpler and more efficient equilibrium techniques to study nonequilibrium systems.

\vskip 1,6 cm

{\bf Disclaimer.} This paper is not intended for journal publication.


\end{document}